\documentclass[3p,times,twocolumn]{elsarticle}

\usepackage{ecrc}

\volume{00}

\firstpage{1}

\journalname{Nuclear Physics B Proceedings Supplement}

\runauth{M. Golterman {\it et al.}}

\jid{nppp}

\jnltitlelogo{Nuclear Physics B Proceedings Supplement}

\usepackage{amssymb}

\usepackage[figuresright]{rotating}

\begin{document}

\begin{frontmatter}



\dochead{}

\title{The status of the strong coupling from tau decays in 2016}


\author{Diogo Boito}
\address{Instituto de F{\'\i}sica de S{\~a}o Carlos, Universidade de S{\~a}o Paulo, 
CP 369, 13570-970, S{\~a}o Carlos, SP, Brazil}

\author{Maarten Golterman\fnref{label1}}
\fntext[label1]{Speaker}
\address{Department of Physics and Astronomy,
San Francisco State University, San Francisco, CA 94132, USA}

\author{Kim Maltman}
\address{Department of Mathematics and Statistics,
York University,  Toronto, ON Canada M3J~1P3\\
CSSM, University of Adelaide, Adelaide, SA~5005 Australia}

\author{Santiago Peris}
\address{Department of Physics and IFAE-BIST, Universitat Aut\`onoma de Barcelona,
E-08193 Bellaterra, Barcelona, Spain}

\begin{abstract}
While the idea of using the operator product expansion (OPE) to extract
the strong coupling from hadronic 
$\tau$ decay data is not new, there is an ongoing controversy over how
to  include quark-hadron 
``duality violations'' ({\it i.e.}, resonance effects) which are not 
described by the OPE. One approach attempts
to suppress duality violations enough that
they might become negligible, but pays the price of an uncontrolled
OPE truncation.   We critically examine a recent analysis using this 
approach and show that it fails to properly account for 
non-perturbative effects, making the resulting determination of the 
strong coupling unreliable. 
In a different approach duality violations are taken into account with a
model, avoiding the OPE truncation.   This second approach
provides a self-consistent
determination of the strong coupling from $\tau$ decays.
\end{abstract}




\end{frontmatter}


\section{Introduction}
\label{intro}
The recently revised ALEPH data \cite{ALEPH13}\footnote{For the original
ALEPH data, see Ref.~\cite{ALEPH}.} for the vector ($V$) and 
axial ($A$) non-strange hadronic spectral functions extracted from $\tau$
decays led to renewed efforts to extract the strong coupling $\alpha_s(m_\tau^2)$ from these data \cite{ALEPH13,alphas14,Pich}.   These updates
employ different methods:  while perturbation theory and the 
operator product expansion (OPE) are central to both, they differ in the
way they treat resonance effects, or, equivalently, violations of 
quark-hadron duality (DVs) \cite{CGP,russians}.    The method employed in Refs.~\cite{ALEPH13,Pich} (the ``truncated-OPE-model'' strategy)
aims to suppress DVs sufficiently to be able to ignore them altogether,
while the method of Ref.~\cite{alphas14} (the ``DV-model'' strategy)
models DVs explicitly.   These
two approaches lead to values for $\alpha_s(m_\tau^2)$ which differ by
about 8\%, while the errors claimed by each method are significantly
smaller than that.

Both approaches start from finite-energy sum rules (FESRs)
\cite{shankar,Braaten88,BNP}, relating
weighted integrals of the experimental spectral function to a representation
of the theory in terms of $\alpha_s(m_\tau^2)$ and non-perturbative 
effects, captured by higher-dimension terms in the OPE, and, in the case
of the DV-model strategy, a representation of DVs.   The truncated-OPE-model
strategy uses weight functions that suppress DVs and restricts the FESRs
to the largest energy available, {\it i.e.}, $m_\tau$, but requires terms in the
OPE up to dimension 16.   As we will see below, this necessitates
an {\it ad hoc} truncation of the OPE in order to be able to fit the data
provided by the spectral integrals, a truncation for which there is no basis
in QCD.   The DV-model strategy models DVs
explicitly, and makes use of the energy dependence of the sum rules,
requiring the OPE only up to dimension 8, with no truncation necessary.
Both strategies need to be tested for consistency, because both make
assumptions in order to handle non-perturbative effects.

Here we provide a critical appraisal of the truncated-OPE strategy,
focussing on the most recent and very extensive application \cite{Pich} of this
strategy to the ALEPH data.   We focus on the analysis of the $V+A$ non-strange spectral data, as these have been advertized as being least 
affected by duality violations, the neglect of which is central to this
strategy.   We find that this strategy is fundamentally flawed, as a number
of tests show unambiguously that the value for $\alpha_s(m_\tau^2)$
through this strategy is afflicted by uncontrolled systematic errors
related to problems with its treatment of non-perturbative physics.

This brief writeup summarizes only some of the highlights of our analysis.
An extensive account can be found in Ref.~\cite{alphas16}.   For a 
detailed account of the DV-model strategy as applied to the 
determination of $\alpha_s(m_\tau^2)$ from the ALEPH data, including a number of 
self-consistency tests of the assumptions underlying this approach,
we refer to Ref.~\cite{alphas14}.   For the application of the same
strategy to the OPAL data \cite{OPAL} we refer to Refs.~\cite{alphas1,alphas2}.
We note that Ref.~\cite{alphas16}, as a by-product, provides further
support for the DV-model strategy as well.

\section{Theory}
\label{theory}
The extraction of $\alpha_s(m_\tau^2)$ from the non-strange $V$ or $A$ 
spectral functions $\rho^{(1+0)}_{V/A}(s)$ starts from the sum rule
\begin{eqnarray}
\label{sumrule}
&&\hspace{-1cm}\frac{1}{s_0}\int_0^{s_0}ds\,w(s/s_0)\,\rho^{(1+0)}_{V/A}(s)= \\
&&
-\frac{1}{2\pi is_0}\oint_{|s|=s_0}
ds\,w(s/s_0)\,\Pi^{(1+0)}_{{\rm OPE},V/A}(s)\nonumber\\
&&-\frac{1}{s_0}\,
\int_{s_0}^\infty ds\,w(s/s_0)\,\frac{1}{\pi}\,\mbox{Im}\,
\Delta_{V/A}(s)\ ,\nonumber
\end{eqnarray}
in which $\Pi^{(1+0)}_{{\rm OPE},V/A}(s)$ is the OPE approximation to the
exact $V$ or $A$ current two-point function, given by
\begin{equation}
\label{OPE}
\Pi^{(1+0)}_{\rm OPE}(s)=\sum_{k=0}^\infty \frac{C_{2k}(s)}{(-s)^{k}}\ ,
\end{equation}
$\Delta_{V/A}(s)$ represents the part not captured by the OPE, the
so-called ``duality violations'' (DVs), and $w(x=s/s_0)$ is a polynomial weight.
(For a derivation of the sum rule, and the
precise meaning of all quantities, see Ref.~\cite{alphas16}.)
An important observation for the analysis to follow is that a monomial
of degree $n$ in the weight $w(s/s_0)$ picks out the term of order $2k=2(n+1)$ in 
the OPE, because
\begin{equation}
\label{degreecond}
\hspace{-0.7cm}\frac{1}{2\pi is_0}\oint_{|s|=s_0}\!\!\!\!\! ds\,\left(\frac{s}{s_0}\right)^n
\frac{C_{2k}}{(-s)^k} =(-1)^{n+1}\frac{C_{2(n+1)}}{s_0^{n+1}}\,\delta_{k,n+1}\ .
\end{equation}
The $k=0$ term in the OPE represents the mass-independent, purely
perturbative contribution, which is available to $O(\alpha_s^4)$ \cite{PT}.
In the literature, two different resummation schemes, ``fixed-order'' (FO or FOPT) and
``contour-improved'' (CI or CIPT) \cite{CIPT} have been employed.  Following
Ref.~\cite{Pich}, we have considered both.

\section{The truncated-OPE strategy}
\label{truncated}
The analysis of Ref.~\cite{Pich} starts with the weights \cite{DP1992}
\begin{equation}
\label{ALEPH}
w_{k\ell}(x)=(1-x)^{k+2}x^\ell(1+2x)\ ,
\end{equation}
where $(k\ell)\in\{(00),(10),(11),(12),(13)\}$, and takes $s_0=m_\tau^2$ in the 
sum rule (\ref{sumrule}).   Since $C_2$ is negligibly small, the
parameters in the fit to these 5 weighted integrals 
would be $\alpha_s$ and $C_{4,\dots,16}$,
because $w_{13}(x)$ has degree 7 ({\it cf.}\ Eq.~(\ref{degreecond})).
Thus, in order to make a fit possible, one chooses $C_{2k\ge 10}=0$.   
The hope is that the suppression
of the vicinity of the contribution near the real axis at $s=s_0$ by the double or
triple zero in $w_{k\ell}(s/s_0)$ is sufficient to ignore the DV term in Eq.~(\ref{sumrule}).
Ref.~\cite{Pich} carried out tests using other weights, advocating the
``optimal'' weights
\begin{equation}
\label{optimal}
w^{\rm opt}_{2,n}=1-(n+2)x^{n+1}+(n+1)x^{n+2}
\end{equation}
in particular.  In Tables~1 and 2 we reproduce these two fits, for
both FO and CI.
\begin{table}[h!]
\hspace{0cm}\begin{tabular}{|c|c|c|c|c|}
\hline
& $\alpha_s(m_\tau^2)$ & $C_{4}$ (GeV$^4$) & $C_{6}$ (GeV$^6$)& $C_{8}$ (GeV$^8$) \\
\hline
FO & 0.316(3) & -0.0006(3) & 0.0012(3) & -0.0008(3) \\
\hline
CI & 0.336(4) & -0.0026(4)  & 0.0009(3) & -0.0010(4) \\
\hline
\end{tabular}
\caption{{\it Reproduction of the $V+A$ fits of Table 1 of 
Ref.~\cite{Pich}, based on the weights (\ref{ALEPH}).
By assumption, $C_{10}=C_{12}=C_{14}=C_{16}=0$.
Errors are statistical only.}}
\end{table}%
\begin{table}[h!]
\hspace{0cm}\begin{tabular}{|c|c|c|c|c|}
\hline
& $\alpha_s(m_\tau^2)$  & $C_{6}$ (GeV$^6$)& $C_{8}$ (GeV$^8$) & $C_{10}$ (GeV$^{10}$) \\
\hline
FO & 0.317(3)  & 0.0014(4) & -0.0010(5) & 0.0004(3) \\
\hline
CI & 0.336(4)   & 0.0010(4) & -0.0011(5) & 0.0003(3) \\
\hline
\end{tabular}
\caption{{\it Reproduction of the $V+A$ fits of Table 7 of 
Ref.~\cite{Pich}, based on the five ``optimal'' weights (\ref{optimal})
with $n=1,\dots,5$. By assumption, $C_{12}=C_{14}=C_{16}=0$.
Errors are statistical only.}}
\end{table}%

The OPE truncation is not based on QCD, and one might equally
well pick another rather arbitrary, but reasonable set of values, such as
\begin{eqnarray}
\label{OPEMG}
C_{10}&=&-0.0832\ \mbox{GeV}^{10}\ ,\\
C_{12}&=&0.161\ \mbox{GeV}^{12}\ ,\nonumber\\
C_{14}&=&-0.17\ \mbox{GeV}^{14}\ ,\nonumber\\
C_{16}&=&-0.55\ \mbox{GeV}^{16}\ .\nonumber
\end{eqnarray}
Using these values instead of $C_{10-16}=0$ in the fits leads to the results
of Tables~3 and 4.   
\begin{table}[h!]
\hspace{0cm}\begin{tabular}{|c|c|c|c|c|}
\hline
& $\alpha_s(m_\tau^2)$ & $C_{4}$ (GeV$^4$) & $C_{6}$ (GeV$^6$)& $C_{8}$ (GeV$^8$) \\
\hline
FO & 0.295(3) & 0.0043(3) & -0.0128(3) & 0.0355(3)  \\
\hline
CI & 0.308(4) & 0.0031(3)  & -0.0129(3) & 0.0354(3)  \\
\hline
\end{tabular}
\caption{{\it Fits as in Table~1,
but with $C_{10}$, $C_{12}$, $C_{14}$ and $C_{16}$ as given in
Eq.~(\ref{OPEMG}).
Errors are statistical only.}}
\end{table}%
\begin{table}[h!]
\hspace{0cm}\begin{tabular}{|c|c|c|c|c|}
\hline
& $\alpha_s(m_\tau^2)$  & $C_{6}$ (GeV$^6$)& $C_{8}$ (GeV$^8$) & $C_{10}$ (GeV$^{10}$) \\
\hline
FO & 0.295(4)  & -0.0130(4) & 0.0356(5) & -0.0836(3)  \\
\hline
CI & 0.308(5)   & -0.0130(4) & 0.0355(5) & -0.0836(3)  \\
\hline
\end{tabular}
\caption{{\it Fits as in Table~2,
but with $C_{12}$, $C_{14}$ and $C_{16}$ as given in
Eq.~(\ref{OPEMG}).
Errors are statistical only.}}
\end{table}%
All fits shown in these tables are good fits \cite{alphas16}, as good, or
better, in fact, than those of Tables~1 and 2, but the 
alternative results of Tables~3 and 4 are not compatible with those of
Tables~1 and 2.   Clearly, the truncated-OPE strategy allows several,
significantly different solutions.   Which of 
these solutions gets selected by the fits is determined by the arbitrary choice made on the putative values of the higher-order $C_{2k}$. 

\section{Fake data test}
\label{fake}
In order to investigate this problem in more detail, we applied the 
truncated-OPE strategy to a constructed ``fake'' data set, in which 
the value of $\alpha_s$ and the size of the DVs is known, with central
values generated from a model that represents the real-world data
very well, and using real-world covariances \cite{alphas16}.   The 
model-solution for the fit parameters of Tables~1 and 2 is given by
\begin{eqnarray}
\label{modelvalues}
\alpha_s(m_\tau^2)&=&0.312\ ,\\
C_4&=&0.0027~\mbox{GeV}^4\ ,\nonumber\\
C_6&=&-0.013~\mbox{GeV}^6\ ,\nonumber\\
C_8&=&0.035~\mbox{GeV}^8\ ,\nonumber\\
C_{10}&=&-0.083~\mbox{GeV}^{10}\ .\nonumber
\end{eqnarray}
We then apply the fits of the truncated-OPE strategy to these fake data,
and find the results shown in Tables~5 and 6.   Since the fake-data model
was constructed using CIPT, these tables only show CIPT fits.
(A similar test can be carried out using FOPT.)
\begin{table}[h!]
\hspace{0cm}\begin{tabular}{|c|c|c|c|}
\hline
$\alpha_s(m_\tau^2)$ & $C_{4}$ (GeV$^4$) & $C_{6}$ (GeV$^6$)& $C_{8}$ (GeV$^8$) \\
\hline
0.334(4) & -0.0023(4)  & 0.0007(3) & -0.0008(4)  \\
\hline
\end{tabular}
\caption{{\it CIPT fits employing the truncated-OPE strategy 
on the fake data, based on the weights~(\ref{ALEPH}).
By assumption, $C_{10}=C_{12}=C_{14}=C_{16}=0$.
Errors are statistical only.}}
\end{table}%
\begin{table}[h!]
\hspace{0cm}\begin{tabular}{|c|c|c|c|}
\hline
 $\alpha_s(m_\tau^2)$  & $C_{6}$ (GeV$^6$)& $C_{8}$ (GeV$^8$) & $C_{10}$ (GeV$^{10}$) \\
\hline
 0.334(4)   & 0.0008(4) & -0.0008(5) & 0.0001(3)  \\
\hline
\end{tabular}
\caption{{\it CIPT fits employing the truncated-OPE strategy 
on the fake data, based on the ``optimal'' weights~(\ref{optimal}) with 
$m=1$ and $n=1,\dots\,5$. By assumption, $C_{12}=C_{14}=C_{16}=0$.
Errors are statistical only.}}
\end{table}%
Even though the results in Tables~5 and 6 are consistent with one another, and this stability might perhaps lead one to believe in the robustness of these results, clearly, the truncated-OPE strategy fails to find to correct solution, and,
in particular, the value of $\alpha_s(m_\tau^2)$ is off by more than 5~$\sigma$.
The values of the OPE coefficients are also far too small.

\section{Duality violations}
\label{DVs}
\begin{figure}[h!]
\begin{center}
\includegraphics*[width=7.5cm]{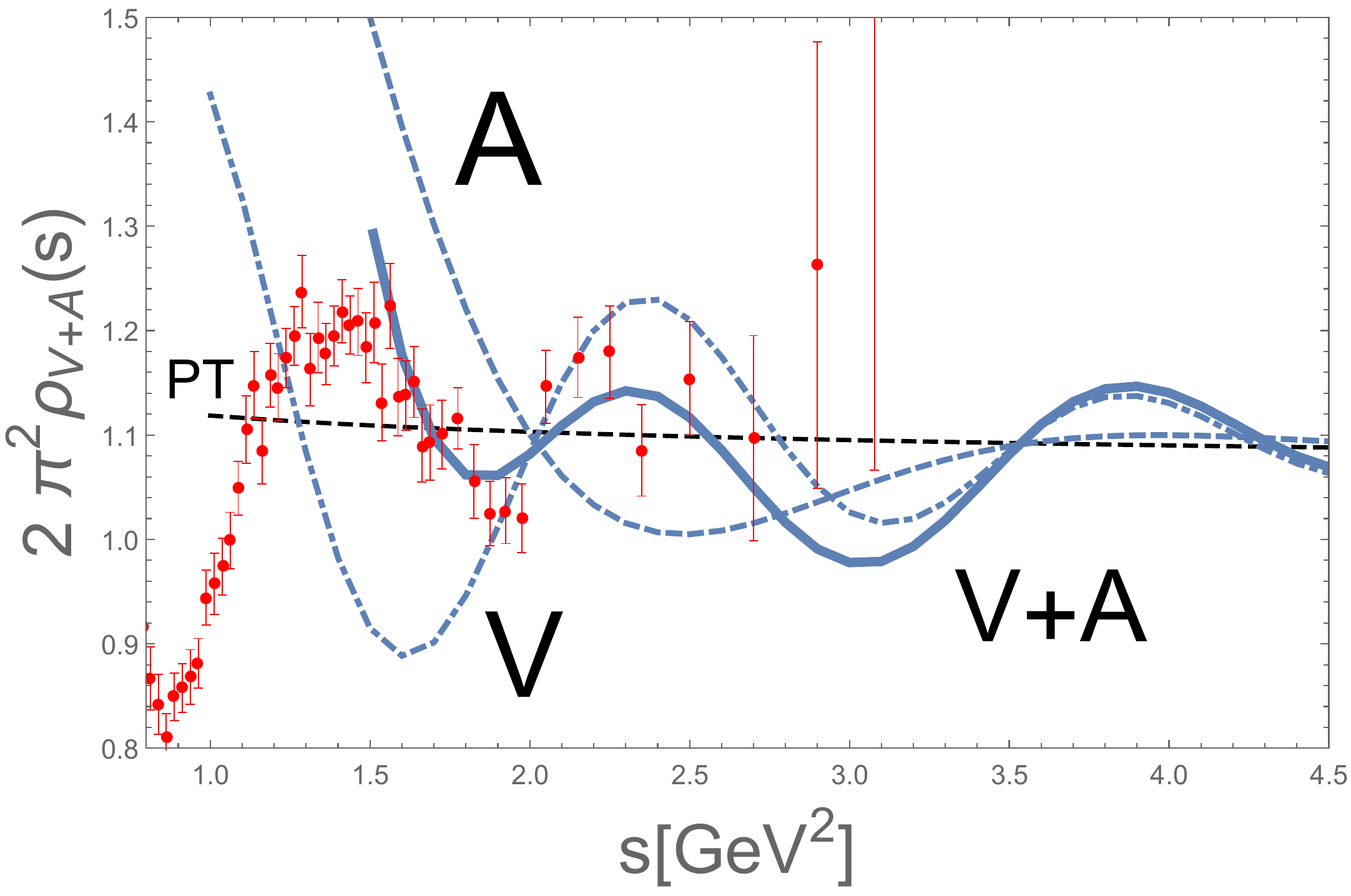}
\end{center}
\begin{quotation}
\vspace*{-4ex}
\caption{\it Blow-up of the large-$s$ region of the $V+A$ non-strange spectral function.
Dashed (black) line: the perturbative (CIPT) representation of
the model. Solid (blue) curve: full model representation, including DVs. 
Dot-dashed (blue) curves: separate $V$ and $A$ parts of the model spectral 
function.}
\end{quotation}
\vspace*{-4ex}
\end{figure}
It is interesting to consider why the truncated-OPE strategy fails
the fake data test. Nominally, it is the uncontrolled truncation
of the OPE that is responsible, but the failure of the OPE truncation
employed is also intimately connected with the presence of DVs 
\cite{alphas14,alphas1}.   In Fig.~1, we show the
large-$s$ region of the ALEPH data for the $V+A$ non-strange spectral
function $\rho_{V+A}$ \cite{ALEPH13}.   The thick blue curve (which is the
sum of the two dot-dashed curves) shows the $V+A$
model underlying the fake data set, and it is perfectly consistent with the 
ALEPH data.   The thin dashed black curve shows the perturbative part of the
spectral function.   Two observations are notable:  first, the dynamics of
QCD, which corresponds to the {\it difference} of the data and the horizontal
line at $2\pi^2\rho_{V+A}(s)=1$ has a large duality-violating contribution
in comparison to perturbation theory, as
evidenced by the clearly visible resonance
oscillations in the data.   Second, the model curve shows a large DV
at $s=m_\tau^2$, larger, in fact, than at any other value of $s$ above
$1.7$~GeV$^2$, even though the DVs are exponentially damped toward
large $s$.   This large effect leads to a significant upward shift in the fitted value of 
$\alpha_s$ away from the exact model value if DVs are entirely ignored, as 
we found in Sec.~4.

\section{Conclusion}
\label{conclusion}
In Ref.~\cite{alphas16} we conducted an extensive study of the truncated-OPE
strategy for obtaining $\alpha_s(m_\tau^2)$ from hadronic $\tau$ decay data,
taking the most recent such analysis of Ref.~\cite{Pich} as our starting point.

A key component of the truncated-OPE strategy is the arbitrary truncation
of the OPE.   In addition, this strategy claims that it is safe to neglect DVs,
employing weighted spectral integrals which suppress their contribution.
We demonstrated in Secs.~3 and 4 that these assumptions cause the
truncated-OPE strategy to fail.   It is not capable of detecting residual
DVs, and, together with the unavoidable, but uncontrolled
truncation of the OPE, it leads to values for
$\alpha_s(m_\tau^2)$ about 8\% too high, even though the error is claimed
to be less than 4\%.   This difference is larger than the long-standing 
difference between FOPT- and CIPT-based results.

Here we only had space to summarize some of our findings, and we refer to
Ref.~\cite{alphas16} for full details, and more discussion comparing the
truncated-OPE and DV-model strategies.   In particular, Ref.~\cite{alphas16}
also contains a refutation of the criticism of the DV-model strategy contained
in Ref.~\cite{Pich}, showing that in fact this ``criticism'' provides further
support for the consistency of the DV-model strategy.

\section*{Acknowledgements}
MG would like to thank the Instituto de 
F{\'\i}sica de S{\~a}o Carlos of the Universidade de S{\~a}o Paulo 
for hospitality and FAPESP for partial support. This material is based 
upon work supported by the U.S. Department of Energy, Office of Science, 
Office of High Energy Physics, under Award Number DE-FG03-92ER40711. 
The work of DB is supported by the S{\~a}o Paulo Research Foundation 
(FAPESP) Grant No. 2015/20689-9 and by CNPq Grant No. 305431/2015-3.
KM is supported by a grant from the Natural Sciences and Engineering Research
Council of Canada.  SP is supported by CICYTFEDER-FPA2014-55613-P, 2014-SGR-1450. 





\begin{thebibliography}{00}


\bibitem{ALEPH13}
  M.~Davier, A.~Hoecker, B.~Malaescu, C.~Z.~Yuan and Z.~Zhang,
  Eur.\ Phys.\ J.\ C {\bf 74}, 2803 (2014)
  [arXiv:1312.1501 [hep-ex]].

\bibitem{ALEPH}
  R.~Barate {\it et al.} [ALEPH Collaboration],
  Eur.\ Phys.\ J.\  C\ {\bf 4}, 409 (1998);
S.~Schael {\it et al.}  [ALEPH Collaboration],
  Phys.\ Rept.\  {\bf 421}, 191 (2005)
  [arXiv:hep-ex/0506072];

\bibitem{alphas14}
  D.~Boito, M.~Golterman, K.~Maltman, J.~Osborne and S.~Peris,
  Phys.\ Rev.\ D {\bf 91}, 034003 (2015)
  [arXiv:1410.3528 [hep-ph]].
  
\bibitem{Pich}
  A.~Pich and A.~Rodr{\'\i}guez-S{\'a}nchez,
  Phys.\ Rev.\ D {\bf 94}, no. 3, 034027 (2016)
  [arXiv:1605.06830 [hep-ph]].

\bibitem{CGP}
  O.~Cat\`a, M.~Golterman, S.~Peris,
  JHEP {\bf 0508}, 076 (2005)
  [hep-ph/0506004];
  Phys.\ Rev.\  D\ {\bf 77}, 093006 (2008)
  [arXiv:0803.0246 [hep-ph]];
  Phys.\ Rev.\  D\ {\bf 79}, 053002 (2009)
  [arXiv:0812.2285 [hep-ph]].

\bibitem{russians}
B.~Blok, M.~A.~Shifman and D.~X.~Zhang,
  Phys.\ Rev.\  D\ {\bf 57}, 2691 (1998)
  [Erratum-ibid.\  D\ {\bf 59}, 019901 (1999)]
  [arXiv:hep-ph/9709333];
  I.~I.~Y.~Bigi, M.~A.~Shifman, N.~Uraltsev, A.~I.~Vainshtein,
  Phys.\ Rev.\  D\ {\bf 59}, 054011 (1999)
  [hep-ph/9805241];
  M.~A.~Shifman,
  [hep-ph/0009131];
  M.~Golterman, S.~Peris, B.~Phily, E.~de Rafael,
  JHEP {\bf 0201}, 024 (2002)
  [hep-ph/0112042].

\bibitem{shankar}
  R.~Shankar,
  Phys.\ Rev.\ D {\bf 15}, 755 (1977);
    R.~G.~Moorhouse, M.~R.~Pennington and G.~G.~Ross, Nucl.\ Phys.\ B\ {\bf 124},
  285 (1977);
    K.~G.~Chetyrkin and N.~V.~Krasnikov,
  Nucl.\ Phys.\ B\ {\bf 119}, 174 (1977);
  K.~G.~Chetyrkin, N.~V.~Krasnikov and A.~N.~Tavkhelidze,
  Phys.\ Lett.\ B\ {\bf 76}, 83 (1978);
  N.~V.~Krasnikov, A.~A.~Pivovarov and N.~N.~Tavkhelidze,
  Z.\ Phys.\ C\ {\bf 19}, 301 (1983);
  E.~G.~Floratos, S.~Narison and E.~de Rafael,
  Nucl.\ Phys.\ B\ {\bf 155}, 115 (1979);
  R.~A.~Bertlmann, G.~Launer and E.~de Rafael,
  Nucl.\ Phys.\ B\ {\bf 250}, 61 (1985).
  
\bibitem{Braaten88}
  E.~Braaten,
  Phys.\ Rev.\ Lett.\  {\bf 60}, 1606 (1988).

\bibitem{BNP}
{E.~Braaten}, {S.~Narison}, and {A.~Pich},
 Nucl.\ Phys.\ B\ {\bf 373} (1992) 581.

\bibitem{alphas16}
  D.~Boito, M.~Golterman, K.~Maltman and S.~Peris,
  arXiv:1611.03457 [hep-ph].

\bibitem{OPAL}
K.~Ackerstaff {\it et al.}  [OPAL Collaboration],
  Eur.\ Phys.\ J.\  C\ {\bf 7} (1999) 571
  [arXiv:hep-ex/9808019].

\bibitem{alphas1}
  D.~Boito, O.~Cat\`a, M.~Golterman, M.~Jamin, K.~Maltman, J.~Osborne and S.~Peris,
  Phys.\ Rev.\ D\ {\bf 84}, 113006 (2011)
  [arXiv:1110.1127 [hep-ph]].

\bibitem{alphas2}
  D.~Boito, M.~Golterman, M.~Jamin, A.~Mahdavi, K.~Maltman, J.~Osborne and S.~Peris,
  Phys.\ Rev.\ D\ {\bf 85}, 093015 (2012)
  [arXiv:1203.3146 [hep-ph]].
  
\bibitem{PT}
P.~A.~Baikov, K.~G.~Chetyrkin and J.~H.~K\"uhn,
  Phys.\ Rev.\ Lett.\  {\bf 101} (2008) 012002
  [arXiv:0801.1821 [hep-ph]].
  
\bibitem{CIPT}
A.~A.~Pivovarov,
  Z.\ Phys.\  C\ {\bf 53}, 461 (1992)
  [Sov.\ J.\ Nucl.\ Phys.\  {\bf 54}, 676 (1991)]
  [Yad.\ Fiz.\  {\bf 54} (1991) 1114]
  [arXiv:hep-ph/0302003];
  F.~Le Diberder and A.~Pich,
  Phys.\ Lett.\  B\ {\bf 286}, 147 (1992).
  
\bibitem{DP1992}
F.~Le Diberder, A.~Pich,
  Phys.\ Lett.\  B\ {\bf 289}, 165 (1992).

\end{thebibliography}



\end{document}